# Gate Tuning of High-Performance InSe-Based Photodetectors Using Graphene Electrodes

Wengang Luo, Yufei Cao, Pingan Hu, Kaiming Cai, Qi Feng, Faguang Yan, Tengfei Yan, Xinhui Zhang and Kaiyou Wang*

Wengang Luo, Yufei Cao, Kaiming Cai, Qi Feng, Faguang Yan, Tengfei Yan, Xinhui Zhang and Kaiyou Wang
SKLSM, Institute of Semiconductors, Chinese Academy of Science, Beijing 100083, China
E-mail: kywang@semi.ac.cn
Pingan Hu,
KLMM, Harbin Institute of Technology, No. 2 Yi Kuang Street, Harbin, 150080, China

**ABSTRACT**

In order to increase the response speed of the InSe-based photodetector with high photoresponsivity, Graphene is used as the transparent electrodes to modify the difference of the work function between the electrodes and the InSe. As expected, the response speed of InSe/graphene photodetectors is down to 120 μs, which is about 40 times faster than that of our InSe/metal device. And it can also be tuned by the back-gate voltage from 310 μs down to 100 μs. With high response speed, the photoresponsivity can reach as high as 60 $AW^{-1}$ simultaneously. Meanwhile the InSe/graphene photodetectors possess a broad spectral range at 400-1000 nm. The design of 2D crystal/graphene electrical contacts could be important for high performance optoelectronic devices.

KEYWORDS: Indium selenide, InSe photodetectors, heterostructure, high response speed, responsivity

## 1. Introduction

Developing novel photodetectors is extremely important in the progress of the optoelectronics field. Among a crowd of various photodetectors, the two dimensional (2D) materials-based photodetectors are very attractive because of their unique dimensional dependent properties. Graphene, being the first prototype of 2D crystals as the channel material in photodetectors, can offer a broad spectral detection and ultrafast sensing due to its linear energy dispersion and high mobility.[1-3] However, the intrinsically weak absorption and small built-in potential in these graphene-based photodetectors have severely limited their photoresponsivity down to $5\times10^{-4}$ $AW^{-1}$ and external quantum efficiency (EQE) to the range of ~0.1-1%.[4] Beyond graphene, novel 2D layered semiconducting materials such as transition-metal dichalcogenides (TMDCs) and several Ⅲ-Ⅵ layered materials have attracted considerably in optoelectronics due to the finite bandgaps.[5-7] Among the III-VI layered materials, the bulk layered InSe has a narrower direct bandgap ($E_g \approx 1.3$ eV) compared with the bandgap of GaS ($E_g \approx 3.05$ eV) and GaSe ($E_g \approx 2.1$ eV),[8,9] which overlaps well with the solar spectrum, thus offering a broader spectral response than GaS- and GaSe-based photodetectors.

Recently the InSe based photodetectors have been reported, one work found that the photoresponsivity of InSe based photodetector can reach 12 $AW^{-1}$, but its response time is only 50 ms.[10] Another work reported a fast response speed (~488 μs), however the responsivity is just $34.7\times10^{-3}$ $AW^{-1}$.[11] Obtaining high photoresponsivity and fast response speed simultaneously is very important for the application of photodetectors. Thus it is very important to increase the response speed of the 2D materials-based photodetectors with high photoresponsivity. The graphene-based artificial heterostructures (stacking of graphene with other layered materials) is used to increase the photoresponsivity of the 2D materials-based photodetectors.[12-17] More recently, the parallel graphene/few-layer InSe conducting heterostructure were used for photodetector, and they found that the photoresponse can be as high as 940 $AW^{-1}$, where the gain photocurrent mainly from the shift of the Dirac point of the graphene layer, Also the response time has not been mentioned in their InSe/G parallel conducting devices.[18] In order to exclude the directly conductance contribution from the Graphene, it is better to design the photodetector InSe device with graphene only working as electrodes. Due to the high mobility of graphene and the close Fermi level between the intrinsic graphene and the InSe, the artificial InSe/graphene heterostructures could dramatically increase the response speed of the InSe-based photodetectors. In this work, the chemical vapor deposition (CVD) growth p-type doping graphene was used as transparent electrodes to generate an artificial heterostructure between graphene and InSe. More details about the p-type doping graphene refer to Supporting Information. A built-in electric field between InSe and graphene will be generated, which will provide a driving force for the separation of photogenerated electrons and holes. The InSe/graphene (InSe/G) heterostructure photodetector shows a broad spectral range at 400-1000 nm and the photoresponsivity can reach as high as 60 $AW^{-1}$. The response time of the InSe/G photodetector is down to 128 μs, which is about 40 times faster than that the reference InSe/metal device. It is noteworthy that the response speed of the InSe/G photodetectors can be effectively tuned by the back-gate voltages.

## 2. Methods

InSe/G photodetectors were fabricated using mechanically exfoliated few-layer InSe nanosheets. For reference, the Ti/Au directly contacted InSe/M photodetector was also fabricated, more information about InSe/M device refer to Supporting Information. For the InSe/G device, a few layer of InSe was exfoliated first on the Si/$SiO_2$ substrate, and then the CVD graphene microstamps were transferred on both ends of the InSe nanosheet, metallic contacts were fabricated on top of the graphene microstamps by using standard electron beam lithography,[19,20] thermal evaporation and lift-off.. The channel length of the InSe/G photodetectors is typically around 16 μm. The schematic illustrations of the InSe/G heterostructure device is presented in Figure 1a. The thickness of the InSe flakes was determined by atomic force microscopy (AFM). The typical thickness used for the sensitive photodetectors in this work is about 30~50 nm. Figure 1d shows the AFM image of the few layered InSe, the thickness is about 33 nm.

**3. Result and discussion**

The crystal structure of InSe is consists of In-Se-Se-In layer as shown in Figure 1b, where each layer has hexagonal structure. The distance between two neighboring layers is 0.84 nm.[21] The high-resolution transmission electron microscopy (HRTEM) image is shown in Figure 1c, confirming the structure of the InSe/G heterostructure. The lattice constant of InSe along *a* or *b* axis is 0.4 nm, which agrees well with the previous results.[22] Two different diffraction patterns are shown in the inset of Figure 1c. Both types of the diffraction patterns present a 6-fold symmetry, indicating the good crystalline quality and also the hexagonal structure for both the graphene (marked by green dash line) and the InSe (marked by red solid line).

Within the wavelength ranging from 400 to 1000 nm, the photoresponsivity of InSe/G

photodetector have very good performance (at light power intensity $P$=0.01 mW $cm^{-2}$, source -drain voltage $V_{ds}$=10 V), as shown in Figure 2a. A well-defined peak is observed at 500 nm, which corresponds to the energy gap of 2.43 eV and is attributed to the optical transitions from $p_x$ and $p_y$-like orbits to the conduction band.[11] The photoresponsivity of the InSe/G photodetector as a function of wavelength slowly decreases from 60 $AW^{-1}$ at $\lambda = 500$ nm to 5.3 $AW^{-1}$ at $\lambda = 1000$ nm.

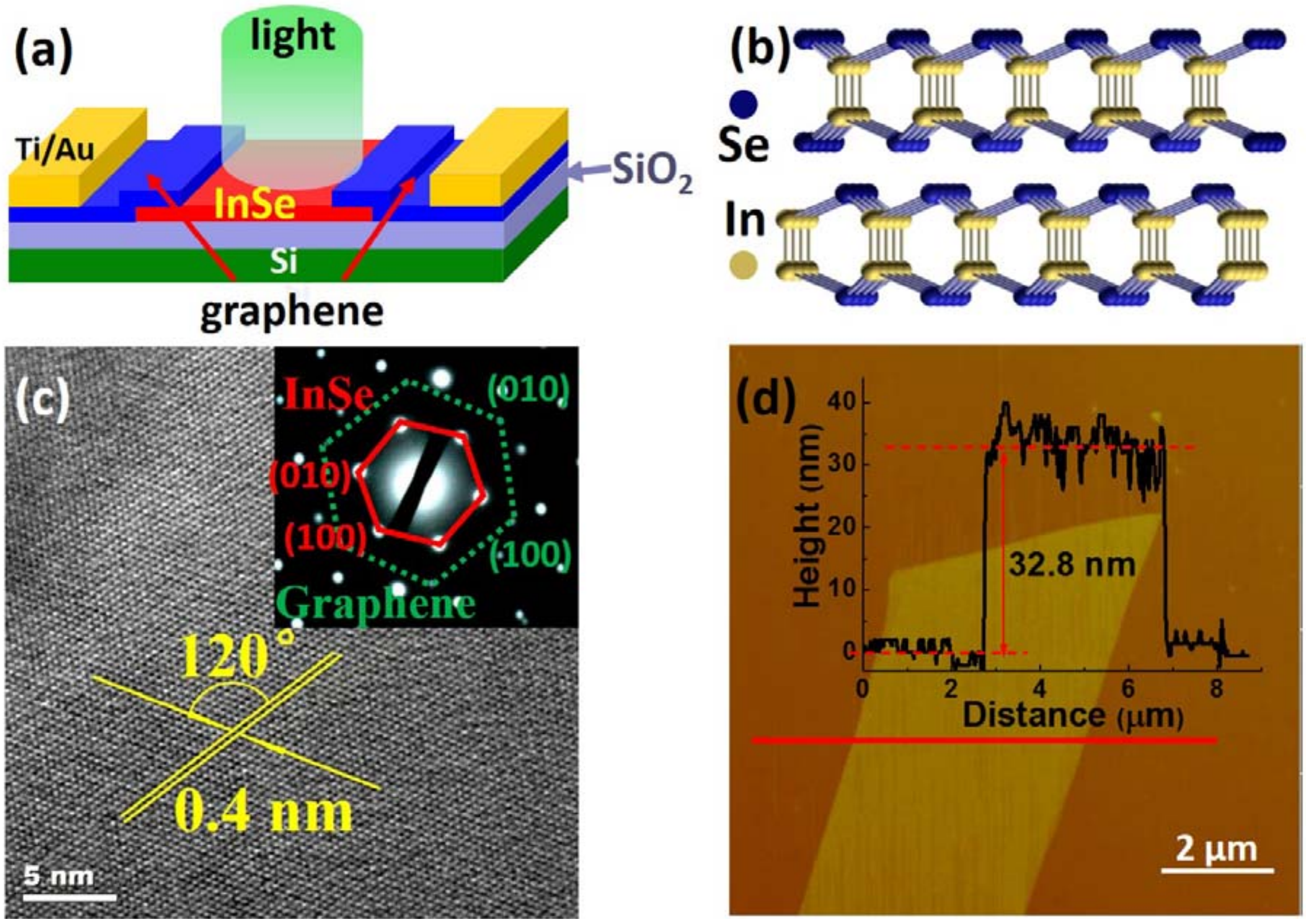

**Figure 1.** (a) A schematic of the InSe/Graphene heterostructure photodetector; (b) The side view of the lattice structure, the distance between two neighboring layer is 0.84 nm; (c) High-resolution TEM image of InSe/G heterostructure. The inset shown two different electron-beam diffraction patterns of InSe (red solid line) and graphene (green dash line); (d) The AFM image of exfoliated InSe flake with a thickness of 32.8 nm.

The EQE defined as the number ratio of electrons flowing out of the device in response to impinging photons, can be expressed as $EQE = hcR/(e\lambda)$, where $h$ is Planck`s constant; $c$ is the light velocity, $R$ is the responsivity, $e$ is the elementary electronic charge, and $\lambda$ is the excitation wavelength. As shown in Figure 2a (the red dots), the EQE has the same wavelength dependence on the photoresponsivity spectrum, where the maximum value of the InSe/G photodetector is ~14850%. The photoresponsivity of InSe/M photodetector has similar wavelength dependence (refer to Supporting Information). But the photoresponsivity of InSe/M photodetector can reach as high as 700 $AW^{-1}$ at $P$=0.01 mW $cm^{-2}$, $V_{ds}$=10 V and $\lambda = 500$ nm. The relatively small photoresponsivity of the InSe/G heterostructure photodetector might be due to two factors. One of the key factors for the Schottky photodetector is the spacing distance between the electrodes. The photoresponsivity exponentially increases with reducing the spacing distance at the fixed source-drain voltage and light intensity, which has been proved in our previous work.[23] Another factor is the interface between the graphene and InSe. The graphene transfer process will bring some defects and charged impurities between the graphene and InSe interface, and this unexpected defects and charged impurities will become recombination center which will restrain some photoinduced charge carriers, and result in a small photocurrent and photoresponsivity.[10] More information about InSe/M photodetector refer to the Supporting Information. To examine the detail performance of devices, the photoresponse at wavelength of 500 nm was chosen for the following studies presented in this work.

To understand how the light intensity and source-drain voltages affect the photodetectors, we first probed the photoresponsivity of the InSe/G photodetectors under various light intensities at $\lambda = 500$ nm. Under the global illumination with light intensity ranging from 0.01 to 2 mWcm$^{-2}$, the illumination intensity dependence of the $I_{ds}$ for the InSe/G photodetectors are shown in Figure 2b. The source-drain current increases rapidly with increasing $V_{ds}$ above 2 V, and it is enhanced with increasing the light intensity. The larger $V_{ds}$ can provide a stronger electric field to decrease the transit time of the photogenerated carriers, and thus reducing the recombination possibility.

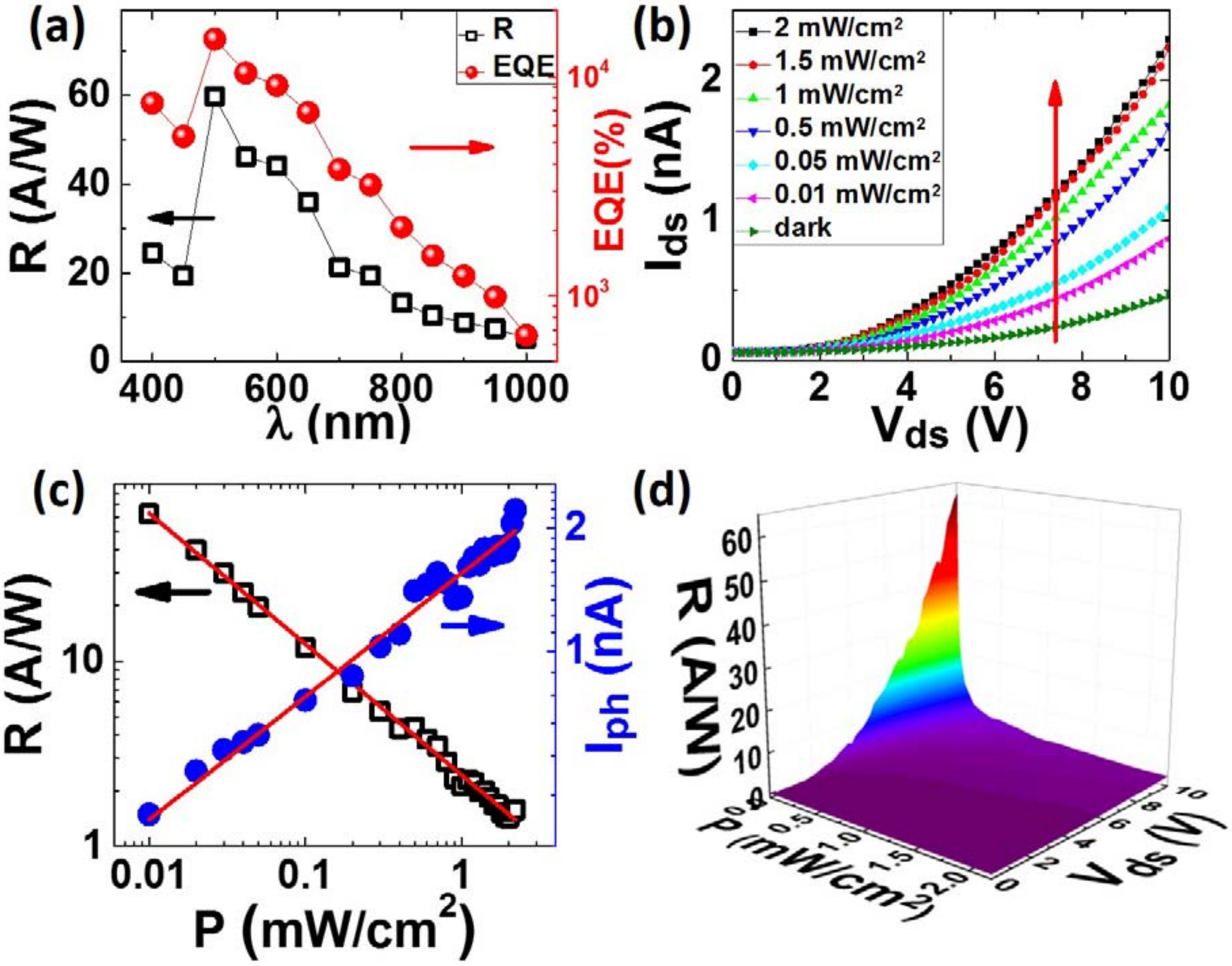

**Figure 2.** (a) The photoresponsivity and EQE of InSe/G photodetector as a function of the illumination wavelength; (b) Typical $I_{ds}$ curves of the InSe/G photodetector with illumination at various excitation intensities (0.01, 0.05, 0.5, 1, 1.5 and 2 mW/cm$^2$) at $V_g$ =0 V, and the dark current of the InSe/G photodetector; (c) Photocurrent (blue solid dots) and responsivity (black open squares) as a function of the illumination intensity at $V_{ds}$ = 10 V and $V_g$ = 0 V,

the red lines are the fitting curves. (d) The 3D responsivity map of the InSe/G photodetector.

Due to the weak light absorption of the graphene, the hole-electron pairs are mainly generated from the InSe under illumination, where the photocurrent originates from the swept electrons and holes in different directions under electric fields. To quantitatively analyze the dependence of illumination intensity upon photoresponse, the photocurrent ($I_{\mathrm{ph}} = I_{\mathrm{light}} - I_{\mathrm{dark}}$) as a function of the light intensity *P* was obtained at fixed source-drain voltage $V_{\mathrm{ds}}$=10 V, as shown in Figure 2c. The photocurrent increases sublinearly following a power law of $I \propto P^{\alpha}$, where $\alpha \approx 0.3$ for InSe/G device while $\alpha \approx 0.45$ for the InSe/M device (refer to Supporting Information). The fitting parameters is much smaller than that of the ideal value of 1.[15] The defects and charged impurities in InSe, InSe/$SiO_2$ and graphene/InSe interface might account for the sublinear power dependence, where more traps could be filled by photoinduced charge carriers with increasing the light intensity, leading to the final saturation of the photocurrent. Similar phenomenon was previously observed in $MoS_2$ photodetectors.[24]

One critical figure-of-merit to determine the performance of the photodetector is the photoresponsivity ($R = I_{\mathrm{ph}} / PS$), which is defined as the ratio of the generated photocurrent ($I_{\mathrm{ph}}$) in response to optical power intensity (*S* is the sample area).[25] The photoresponsivity as a function of illumination power density decreases sublinearly, following a power law of $R \propto P^{-0.71}$. As shown in Figure 2c, the photoresponsivity decreases from 60 to 1.57 $AW^{-1}$ with the illumination intensity increase from 0.01 to 2.2 $mWcm^{-2}$. As shown in the 3D photoresponsivity map of figure 2d, the photoresponsivity can be tuned not only by the illumination intensity but also by the source-drain voltages. The increasing $V_{ds}$ can shorten the carriers` transit time by providing a stronger electric field to govern the photoinduced carries reaching the electrodes, thus reducing the possibility of recombination.

The time-dependent photoresponse of the InSe/G photodetector, under the global illumination with light intensity of 2 $mWcm^{-2}$ at different bias voltages, is shown in Figure 3a. The sensitive, fast and reversible switching between the on and off states allows the device to act as high-quality photodetectors and switchers. The dynamic response to the light illumination for rising and falling in our devices can be expressed by $I(t) = I_0[1 - \exp(-t/\tau_r)]$ and $I(t) = I_0 \exp(-t/\tau_d)$, respectively. Where $\tau_r$ and $\tau_d$ are the time constants for the rising and decay response. Comparing the fitted results between the InSe/G heterostructure and InSe/M devices, the rising and falling time of the InSe/G device are much faster than that of the InSe/M device. The photocurrent of InSe/M device rises dramatically within 4.8 ms after the light illumination and the falling time of $\tau_d$ is about 5.6 ms, as shown in Figure 3b inset. But for the InSe/G photodetector, the rising and falling time decrease dramatically to 120 μs and 220 μs (Figure 3b), respectively. To our knowledge, that is superior to all 2D material-based photodetector (except graphene photodetector). Notably, as shown in Figure 3a and Figure S5a, the falling edge of the photocurrents exhibits two-step relaxation with a rapid fall in the first step ($\tau_{d1} < 10$ ms) and a slow fall in the second step ($\tau_{d2} > 1$ s). This phenomenon was also observed in previous studies.[10,26,27] By fitting the second decaying stage, almost the same long decay time constant (5.83s for InSe/G and 5.65s for InSe/M) was obtained (Figure S5b). The two different time constants in the decaying stage imply the existence of various traps in the sample. And the relatively long decay time constant can be attributed to the inherent trap states in InSe nanosheet. A list of the performance metrics for comparison among the recently developed 2D material-based photodetector is provided in Table S1 of the Supporting Information. The much fast response of the InSe-based photodetector suggests that using graphene as transparent electrodes can effectively improve the InSe photodetectors.

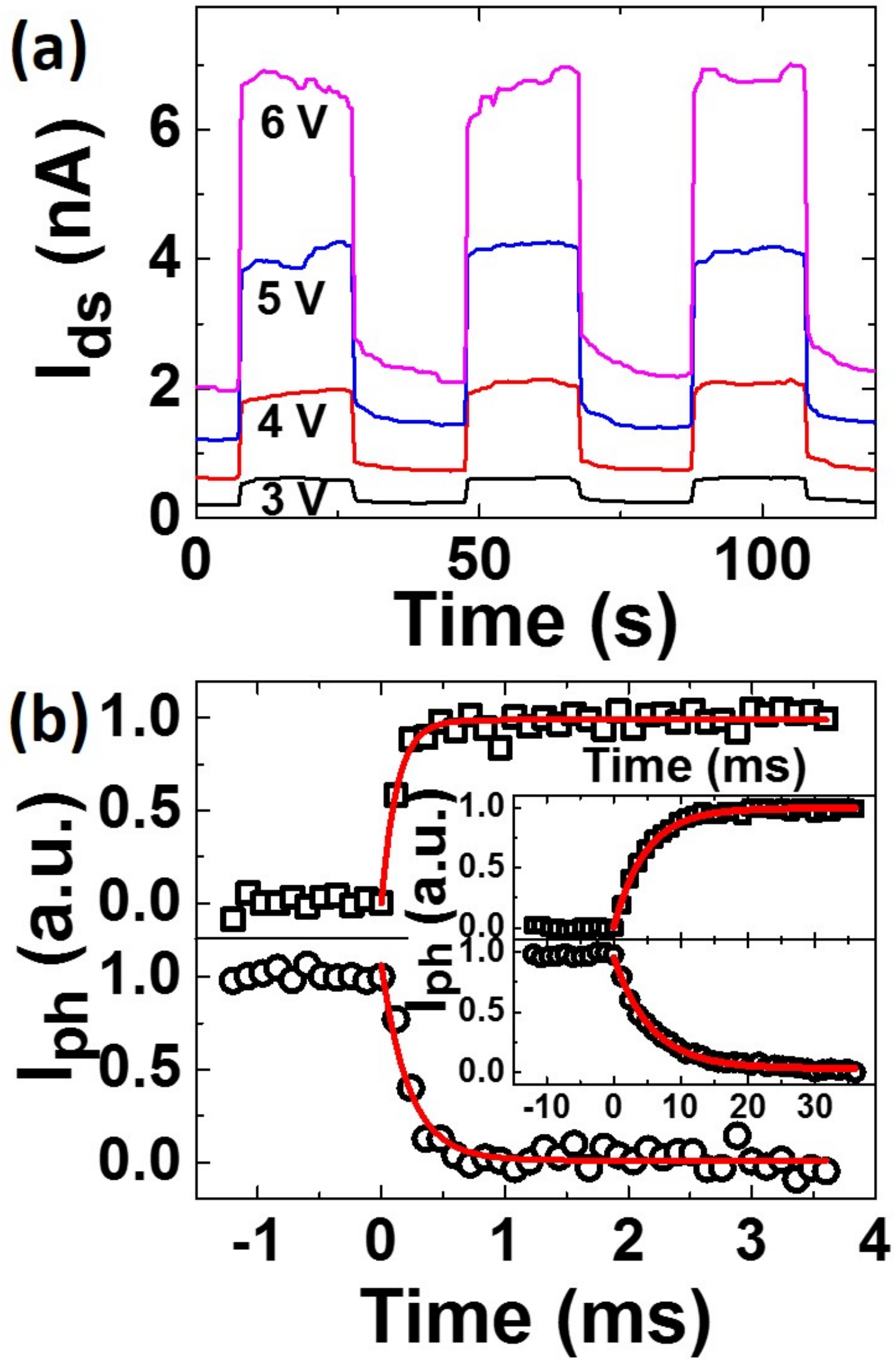

**Figure 3.** (a) The InSe/G photodetectors exhibit long-time stability in photoswitching; (b) The rising and falling time of the InSe/G photodetectors, inset shows the rising and falling time of the reference InSe/M photodetector. Red lines are the fitting curves.

In order to investigate the influence of the p-type doping graphene on the InSe photodetector, we explored the dependence of the current profile on the back-gate voltages ($V_g$) in the InSe/G photodetector. As shown in Figure 4a, at fixed $V_{ds} = 10$ V and light intensity $P$=2.5 mW cm$^{-2}$, the photocurrent firstly decreases with sweeping the $V_g$ from -80 to 40 V, while it increases with increasing $V_g$ further from 40 to 80 V. As shown in Figure 4a up-inset, the dark current also decreases with sweeping the $V_g$ from -80 to 0 V, and it increases when sweeping $V_g$ from 0 to 40 V, but it decreases with increasing Vg further from 40 to 80 V. This dependence of the dark current on the back-gate voltages results from the difference between the source-drain voltage and the Schottky barriers between graphene and InSe nanosheet. (More information about the dark current at different back-gate voltages refer to Supporting Information section 4). Interestingly, the rising time exhibits the same dependence on $V_g$ as that of $I_{ph}$, which can be tuned from 310 to 100 μs by the $V_g$, as shown in Figure 4b. The same dependence of the photocurrent and the response time on the back-gate voltages suggests they have the same physical origin. The back-gate voltages can tune the Fermi level of graphene and InSe, which will change the Schottky barrier between graphene and InSe. When graphene and InSe are in contact, the Fermi levels must coincide at the interface. The work function of intrinsic graphene is about 4.56 eV,[28] and the Fermi level of InSe ($E_f$(InSe)) is about 4.45 eV.[29] Before applying a back-gate voltage, there is an initial built-in potential between the p type graphene and n type InSe interface. Under illumination, the photogenerated carriers in the InSe/G device can immediately move to the graphene layer due to the built-in electric field and applied electrostatic field. The much smaller Schottky barrier of the InSe/G device and the high mobility of graphene result in much faster response speed to that of the InSe/M devices.

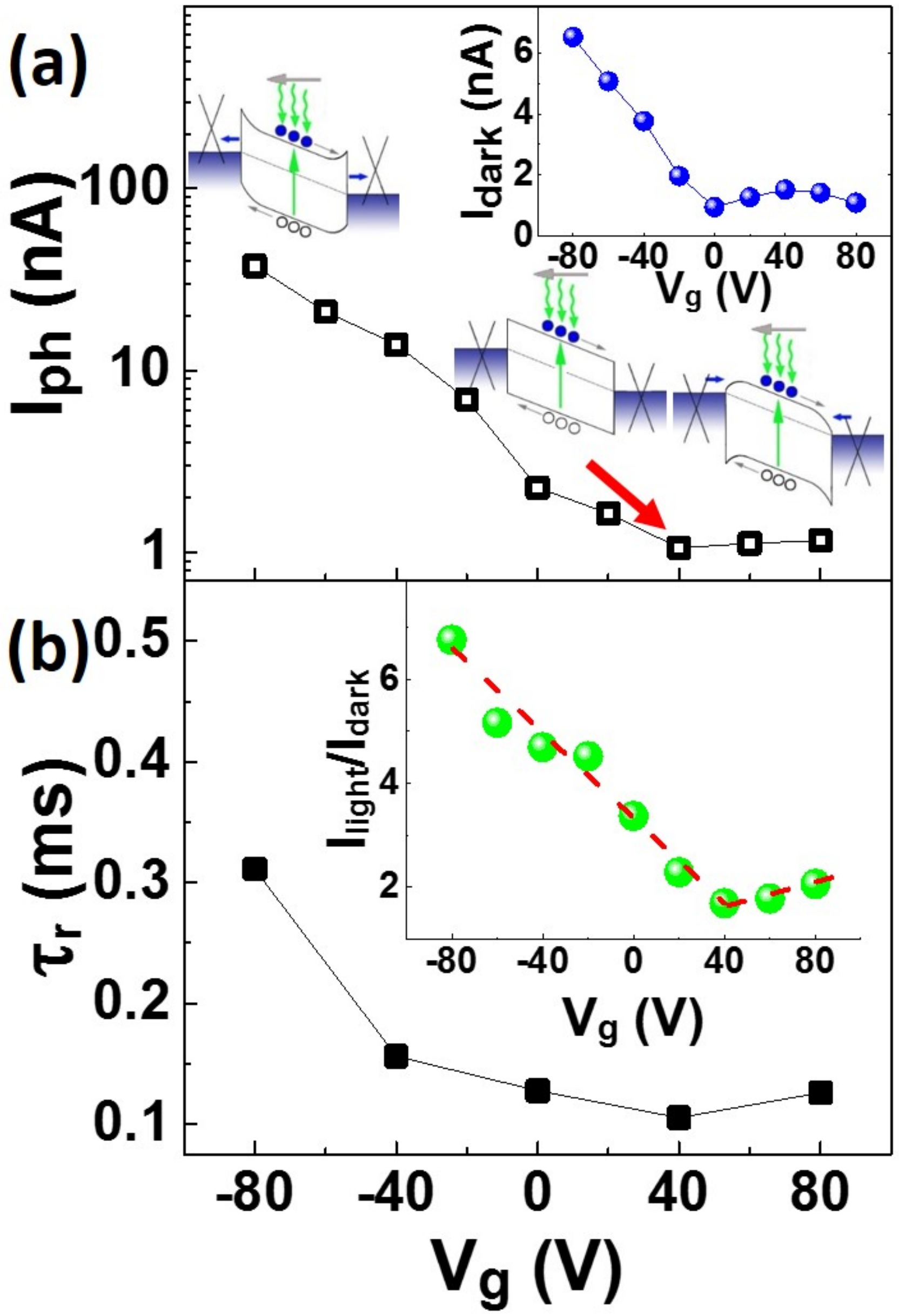

**Figure 4.** (a) Photocurrent of the InSe/G photodetector at different $V_g$, with $V_{ds}$=10 V and $W$=2.5 mWcm$^{-2}$. Middle-inset shows the energy band diagrams of the InSe/G photodetector under illumination at different $V_g$. Up-inset presents the dark current of InSe/G dependence on $V_g$ ($V_{ds}$=10 V); (b) Response time at different $V_g$, with $V_{ds}$=10 V and $W$=2.5 mWcm$^{-2}$. (Inset) Photoresponse ratio ($I_{lihght}/I_{dark}$) as a function of $V_g$.

The mechanism of the back-gate voltages tuning the photocurrent and response speed can be divided into three situations where the Fermi level of graphene ($E_f$(G)) is under, equal and above to that of InSe. When applying a negative back-gate voltage ($E_f$(G) < $E_f$(InSe)) (left band diagram in Figure 4a), the down-shifted Fermi level of InSe results in a larger energy barrier between the graphene and the conduction band of InSe. Meanwhile the Fermi level of graphene is shift down, lead to more p doping, which will increase the built-in potential between the graphene and InSe. Therefore, upon illumination, more photogenerated carriers can be separated by the bigger built-in potential, which will result in a relatively large photocurrent. But the large Schottky barrier will increase the capacitance and the resistance between the graphene and InSe, which will decrease the response speed. With increasing the back-gate voltages from negative to 40 V, the Schottky barrier between the graphene and InSe decreases, which results in the increase of the response speed. When $V_g$=40 V, the Fermi level of graphene will equal to that of InSe (middle band diagram in Figure 4a), there will be no built-in potential between InSe and graphene. When the photocurrent decreases to the minimum value, the response speed increases to the maximum value. Continuing to increase the back-gate voltage further, the Fermi level of graphene will be higher than that of the InSe, so that the opposite Schottky barrier and built-in potential between graphene and InSe are formed again, as shown in the right side band diagram of Figure 4a. Similar to the first case, the Schottky barrier and built-in potential will increase the photocurrent and decrease the response speed. Meanwhile the Schottky barrier will restrict the tunneling and the thermionic currents. The measured $I_{ds}$-$V_{ds}$ curves of InSe/G at different back-gate voltages proved the correctness of our mechanism (refer to Supporting Information section 4). As shown in Figure S6a, the non-linear $I_{ds}$-$V_{ds}$ curves, at $V_g$= -80 V, -40 V, 0 V and +80 V, identify the non-Ohmic contacts. Meanwhile, when $V_g$ equals to 40 V, the output curve is almost linear (Figure S6a inset), indicating the absence of built-in potential thus exhibiting the Ohmic behavior. The photoresponse ratio ($I_{light}$/$I_{dark}$, as shown in the inset of Figure 4b) has the similar dependence of $V_g$ with that of $I_{ph}$, and this dependence demonstrates that the Schottky barrier will restrict the tunneling and the thermionic current.

## 4. Conclusion

In summary, we used graphene as transparent electrodes to fabricate InSe/G photodetectors on $SiO_2$/Si substrate and decreased its response time down to 120 μs. And the response time can be tuned from 310 μs down to 100 μs by the back-gate voltage, which is about 40 times faster than that of our InSe/M device. Meanwhile the InSe/G photodetectors have high responsivities over a broad spectral range at 400-1000 nm. The high responsivity ($R$=60 $AW^{-1}$) and broad spectral response (from visible to near-infrared) are important for wide spectral photodetectors. Our work suggests that the response speed of 2D material photodetector can be improved by using the graphene as electrodes to design heterostructure photodetectors with high photoresponsivity, which could be very important for future integrated optoelectronic applications.

**Supporting Information**

Supporting Information is available from the Wiley Online Library or from the author.

**Acknowledgements**
This work was supported by "973 Program" Nos. 2011CB922201, 2014CB643903, and NSFC Grant Nos. 61225021, 11474276.

Received: ( )
Revised: ( )
Published online: ( )

# Supporting Information

## Electrical Contacts Effect in InSe Nanosheets-based Photodetector with High Response Speed and Photoresponsivity

Wengang Luo, Yufei Cao, Pingan Hu, Kaiming Cai, Qi Feng, Faguang Yan, Tengfei Yan, Xinhui Zhang and Kaiyou Wang*

**Supporting Information Contents**

## 1. The property of CVD graphene

The graphene was grown by CVD method, and then the graphene was transferred onto $Si/SiO_2$ substrate. Figure S1a shows the optical images of the transferred graphene onto $Si/SiO_2$ substrate. The transferred graphene is very uniform in a large area. And the physical properties of the graphene was investigated by the Raman spectra at wavelength 532 nm, as shown in Figure S1b. The Raman spectra of the graphene showed a $I_{2D}/I_G$ ~2 to indicate monolayer graphene.[1]

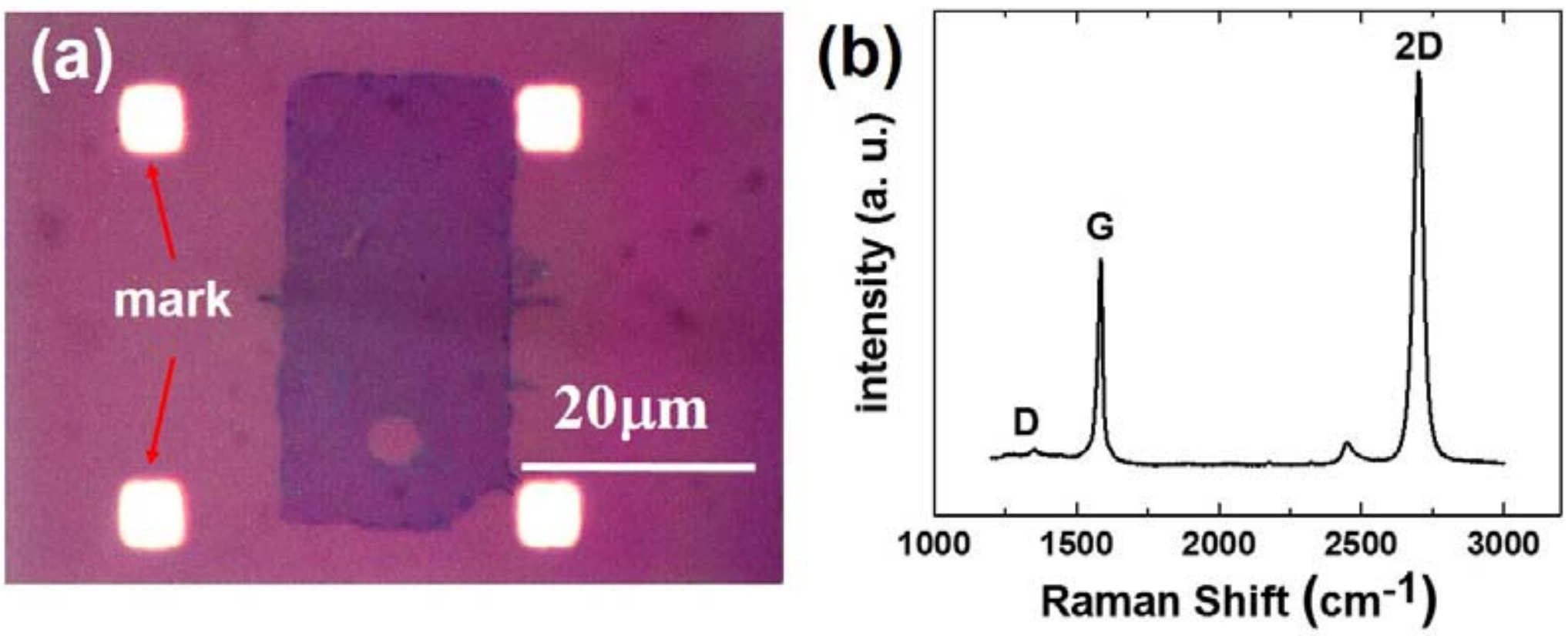

**Figure S1.** (a) The optical images of transferred graphene onto $Si/SiO_2$ substrate. (b) The Raman spectrum of graphene.

Our graphene field effect transistors (FET) show bipolar behavior with scanning the back-gate voltages. The neutral point of the device shown in Figure S2 shifts to 22 V, indicating unintentionally p doping CVD growth graphene of $1.2\times10^{12}$ $cm^{-2}$. The carrier mobility $\mu$ can be calculated from $\mu= (L/WC_gR^2)dR/dV_g$, where $L$=14μm is the channel length, $W$=5μm is the channel width, $V_g$ is back-gate voltage and $C_g$= 115 aF/μm$^2$ is the capacitance between the channel and the back-gate per unit area ($C_g=\varepsilon\varepsilon_0/d$; $\varepsilon$= 3.9; $d$=300 nm). The hole (electron) mobility of the graphene is calculated to be 4300 (1900) $cm^2V^{-1}s^{-1}$, which is good enough for most of the electronic applications.

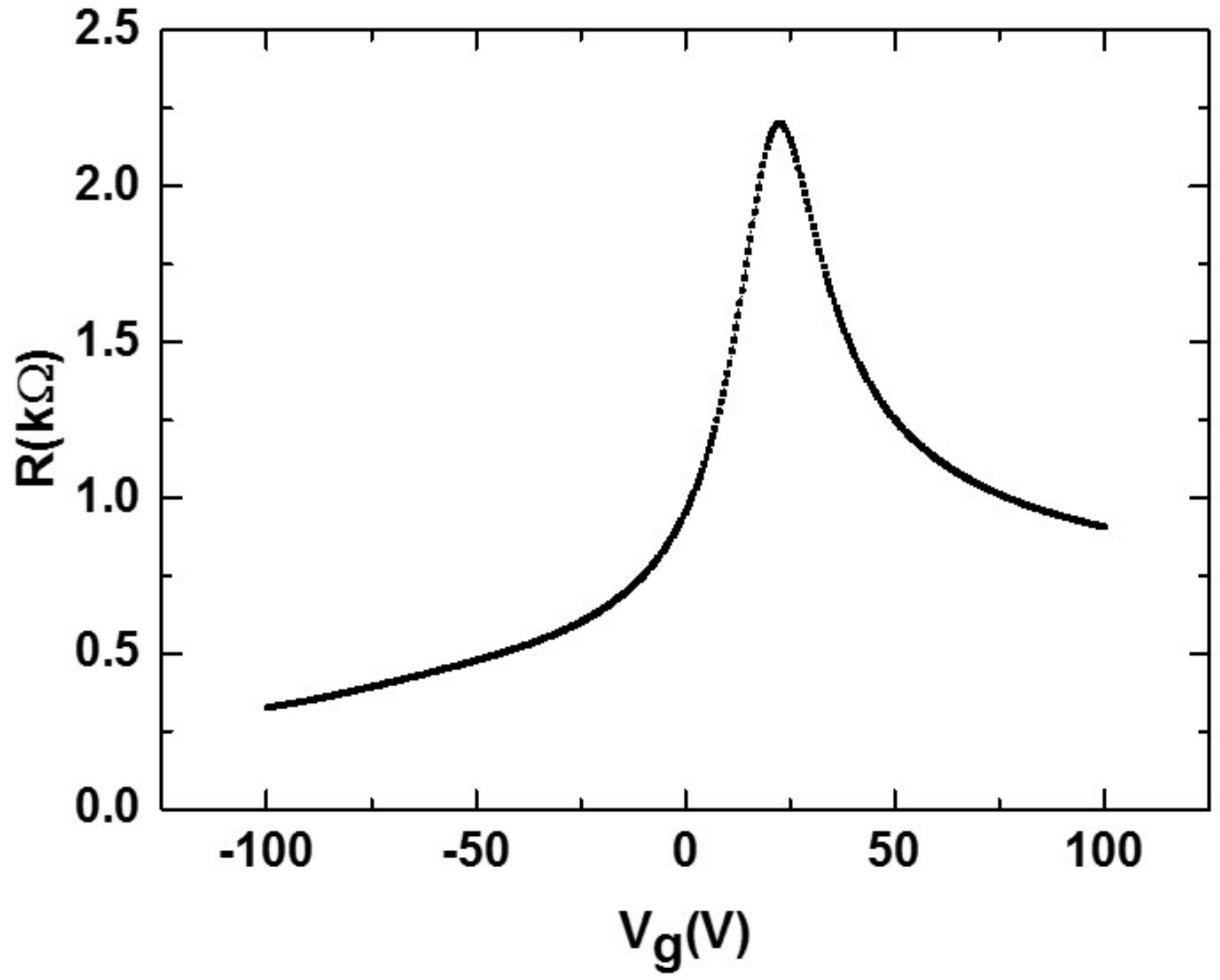

**Figure S2.** The typical sheet resistance vs. back-gate voltages at fixed source drain voltage at 1 mV.

## 2. The property of InSe/M photodetector

For reference, the Ti/Au directly contacted InSe photodetector (InSe/M) was also fabricated. For the InSe/M photodetector, a few layer of InSe was exfoliated first on the $Si/SiO_2$ substrate, and then the metallic contacts (Ti/Au ~ 10/100 nm) were thermally evaporated on top of it. Figure S3 left inset shows the schematic illustrations of the InSe/M photodetector.

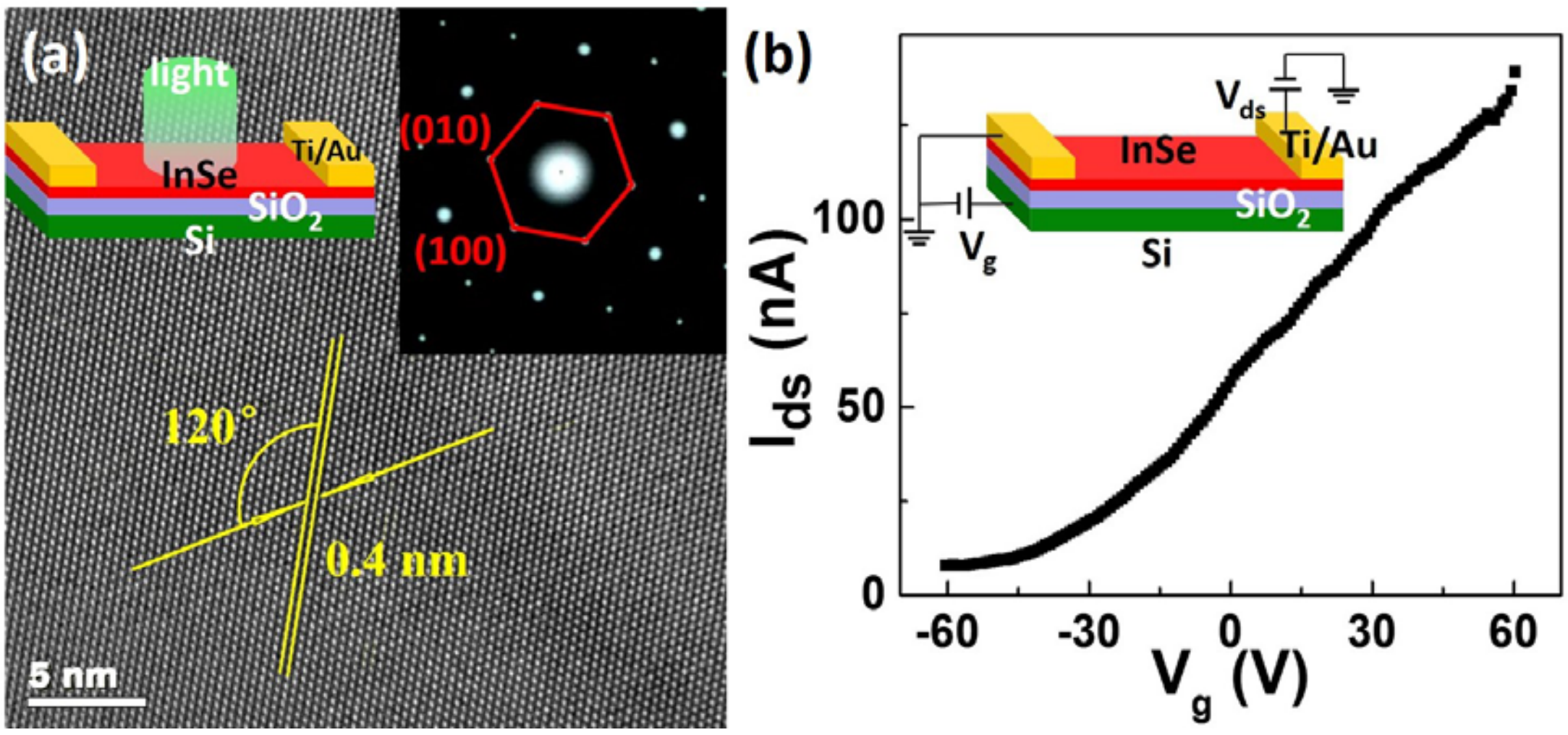

**Figure S3.** (a) High-resolution TEM image of InSe. The left inset shows the schematic of the few-layered InSe photodetector, the right inset shows the different electron-beam diffraction pattern of InSe. (b) The transfer curve of a few-layered InSe FET was measured by scanning $V_g$ from -60 to 60 V at $V_{ds}$=10 V. Top inset presents the schematic illustration of the InSe-FET electrical measurement.

The high-resolution transmission electron microscopy (HRTEM) image of InSe is shown in Figure S3. The lattice constant of InSe along the *a* or *b* axis is 0.4 nm. The diffraction pattern is shown in the right inset of Figure S3. The diffraction pattern of InSe presents a 6-fold symmetry, indicating the good crystalline quality and also the hexagonal structure for the InSe. As shown in Figure S3b, it is "switched off" at the positive back-gate bias, which indicates the n-type of InSe.

As shown in Figure S4a, the InSe/M photodetector has very good performance with the wavelength ranging from 400 to 1000 nm (at light power intensity $P$=0.01 mWcm$^{-2}$, source -drain voltage $V_{ds}$=10 V), At $\lambda = 500$ nm, the photoresponsivity and the EQE of InSe/M photodetector is calculated to be 700 AW$^{-1}$ and ~174000%, respectively, as shown in Figure

S4a (the red dots).

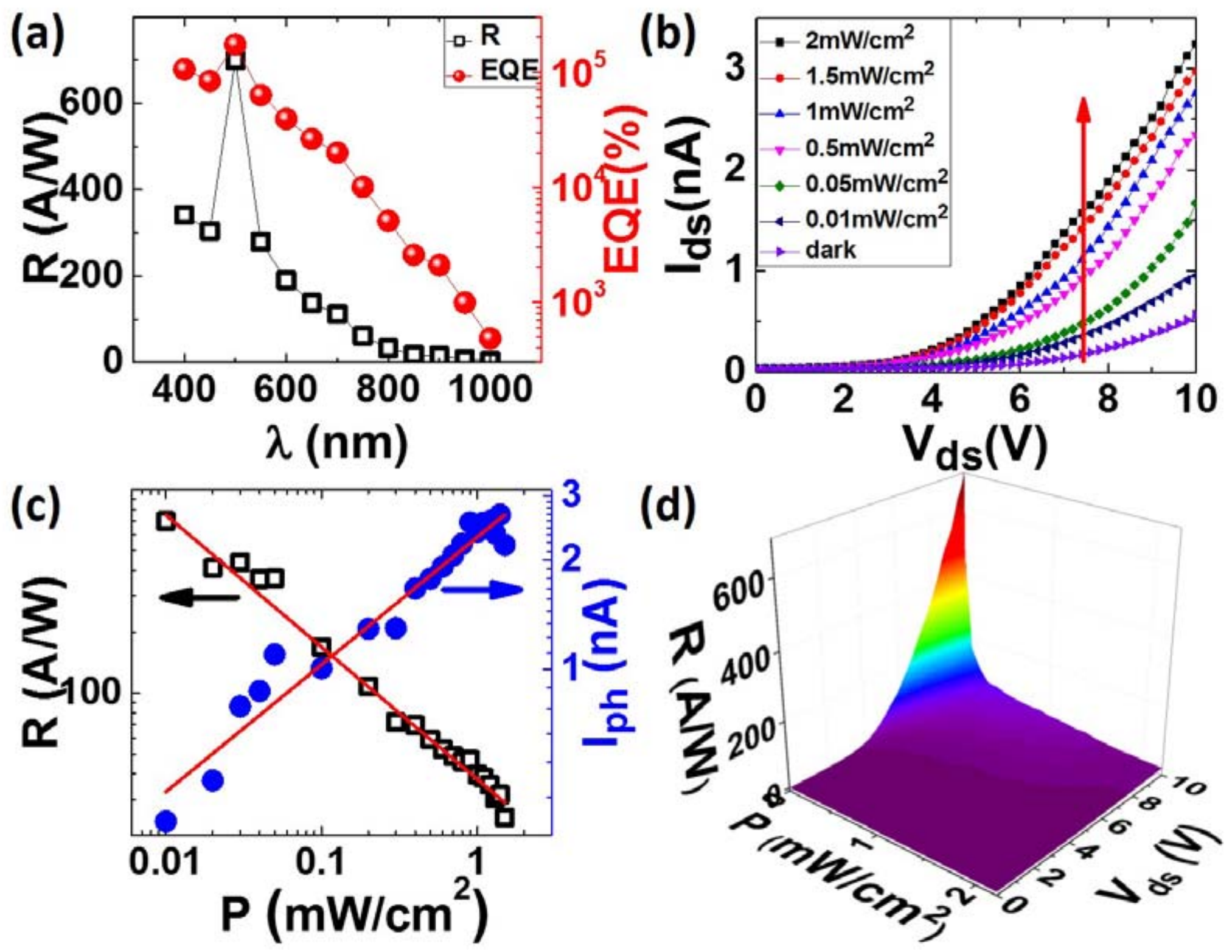

**Figure S4.** (a) The photoresponsivity and EQE of InSe/M photodetector as a function of the illumination wavelength; (b) Typical $I_{ds}$ curves of the InSe/M photodetector with illumination at various excitation intensities (0.01, 0.05, 0.5, 1, 1.5 and 2 mW/cm$^2$) at $V_g$ =0 V, and the dark current of the InSe/M photodetector; (c) Photocurrent (blue solid dots) and responsivity (black open squares) as a function of the illumination intensity at $V_{ds}$ = 10 V and $V_g$ = 0 V, the red lines are the fitting curves. (d) The 3D responsivity map of the InSe/G photodetector.

As shown in Figure S4c. The photocurrent increases sublinearly following a power law of $I \propto P^{\alpha}$, where $\alpha \approx 0.45$ for InSe/G device. Figure S4c presents the photoresponsivity (black open squares) of InSe/M photodetectors as a function of illumination power density of $R \propto P^{-0.55}$. The time-dependent photoresponse of the InSe/M photodetector, under the global illumination with light intensity of 1 $mWcm^{-2}$ at different bias voltages, is shown in Figure S5a.

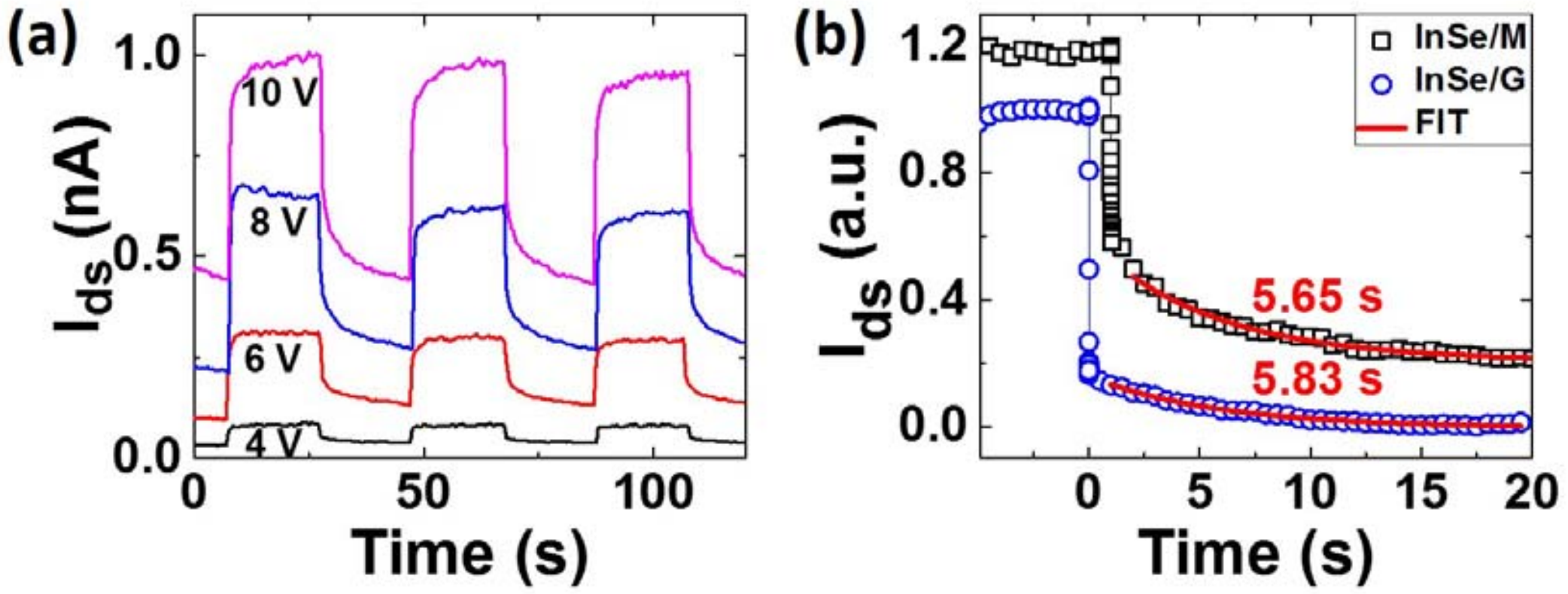

**Figure S5.** (a) The photocurrent as a function of time. (b) The decaying stages of the InSe/M and InSe/G device. The red lines are the fitting line about the long decay part.

### 3. The specific detectivity of InSe/G and InSe/M photodetector

The specific detectivity ($D^*$) is a very important criterion reflecting the photodetector`s sensitivity. Since the primary source of noise to limit $D^*$ is the shot noise from dark current, the specific detectivity can be expressed as $D^* = RS^{1/2} / (2eI_{dark})^{1/2}$,[2] where $R$ is responsivity, $S$ is the area of the phototransistor channel, $e$ is the electron charge, and $I_{dark}$ is the dark current. Under illumination of $\lambda$=500 nm and $P$=0.01 mW $cm^{-2}$ at $V_{ds}$=10 V, the calculated $D^*$ is ~ $2\times10^{13}$ Jones for few-layered InSe/M phototdetector and $D^*$ is ~$2.5\times10^{12}$ Jones for InSe/G heterostructure photodetector, which surpasses to the both $D^*$ values of InGaAs (~$10^{11}$-$10^{12}$ Jones) and $MoS_2$ photodetectors (~$10^{11}$ -$10^{12}$ Jones).[3,4] The difference in D* is mainly due to the different photoresponsivity. The photoresponsivity of the InSe/M (700 $AW^{-1}$) is more than one order of magnitude greater than that of the InSe/G device (60 $AW^{-1}$). The sample size of InSe/G is only three times larger than that of InSe/M device, and both devices also showed similar dark current. The thickness of InSe/M device is about 35 nm, which is similar to that of the InSe/G device (33 nm).

## 4. The $I_{ds}$-$V_{ds}$ curves of InSe/G device

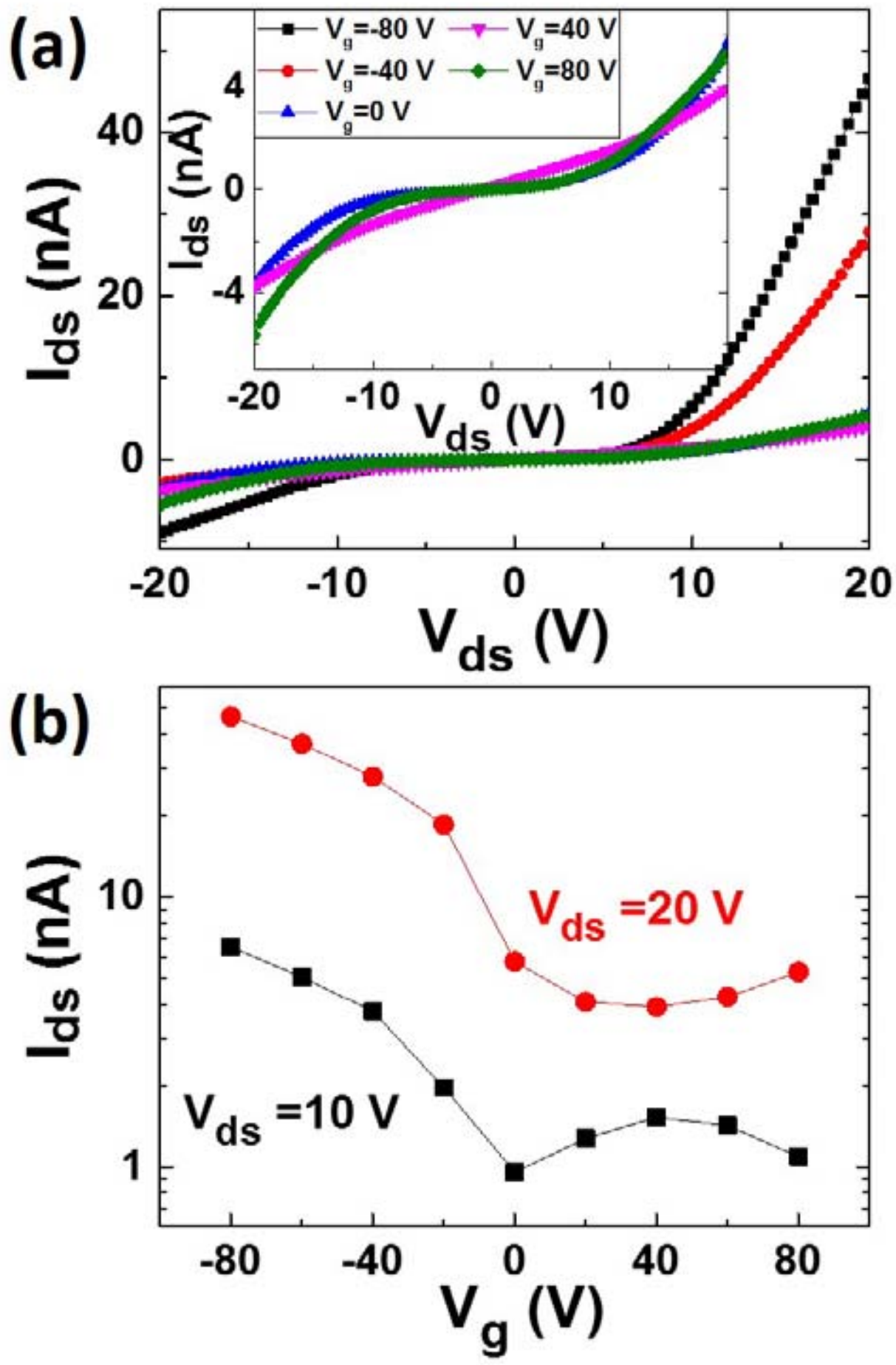

**Figure S6. (a)** The $I_{ds}$-$V_{ds}$ curves at different back-gate voltages (-80, -40, 0, 40 and 80 V.). Inset shows the detail of the output curves at $V_g$=0, 40 and 80 V. (b) The dependence of dark current on $V_g$ with $V_{ds}$ at 20 V ( red solid circles) and 10 V ( black solid squares).

We measured the $I_{ds}$-$V_{ds}$ curves at different back-gate voltages without light illumination. As shown in Figure S6a, the nonlinearity curves at $V_g$=-80, -40, 0 and 80 V indicate the non-Ohmic contact. The asymmetry of the $I_{ds}$-$V_{ds}$ curves when the $V_g$ is smaller than +40 V is due to the generation of p-n junction between graphene and InSe. When $V_g$ equals to 40 V, the output curve is almost linear, indicating the absence of built-in potential thus exhibiting the Ohmic behavior. We also presented the dependence of dark current on $V_g$ at different $V_{ds}$. As shown in Figure S6b, when sweeping $V_g$ from 0 to 80 V, the dark current shows different dependence on $V_g$ at different source drain voltages. The back-gate voltages can effetely tune the Schottky barriers. This different dependence of dark current on $V_g$ at different source drain voltages is originated from the magnitude difference between the source drain voltage and the Schottky barriers.

## 5. Summary of different 2D material photodetectors

Some 2D material based photodetectors show higher responsivity but much slower response time and limitation in the detection range. The combination of broadband (Visible to NIR), sizeable (R=60 AW$^{-1}$) and fast (100 μs) photoresponse makes InSe/G a very promising method for photodetection.

**Table S1**. Comparison of figures-of-merit for photodetectors based on 2D materials

| material (Bandgap in eV) | Measurement condition | Responsivity (A/W) | EQE or IQE (%) | Response time | Spectral window | Ref. |
|---|---|---|---|---|---|---|
| Graphene/InSe/ Graphene heterostructure | 500 nm $V_g = 0V$ $V_{sd} = 10V$ | 60 | 14850 | 120 μs | Visible-NIR | This work |
| InSe/M (1.3, DB) | 500 nm $V_g = 0V$ $V_{sd} = 10V$ | 700 | 174000 | 4.8 ms | Visible-NIR | This work |
| Graphene (gapless) | 1550 nm $V_g = 80V$ | $5\times10^{-4}$ | 6-16 (IQE) | ~ ps | UV-IR | [5] |
| Monolayer $MoS_2$ (1.8, DB) | 561 nm $V_g = -70V$ $V_{sd} = 8V$ | 880 | NA | 600 ms | Visible | [6] |
| Few-layered GaSe (2.1, IB) | 254 nm $V_{sd} = 5V$ | 2.8 | 1367 | 20 ms | UV | [7] |
| Few-layered GaS (3.05, IB) | 254 nm $V_{sd} = 2V$ | 4.2 | 2050 | <30 ms | UV | [8] |
| Few-layered $WS_2$ (1.9, DB) | 458 nm | $21.2\times10^{-6}$ | NA | 5.3 ms | Visible | [9] |
| Few-layered InSe (1.3, DB) | 532 nm $V_{sd} = 3V$ | $34.7\times10^{-3}$ | 8.1 | 488 μs | Visible-NIR | [10] |
| Few-layered InSe on $SiO_2$/Si (1.3, DB) | 450 nm $V_g = 0V$ $V_{sd} = 10V$ | 12.3 | 3389 | 50 ms | Visible-NIR | [11] |
| Graphene/InSe Heterostructure | 532 nm $V_g = 0V$ $V_{sd} = 50mV$ | 940 | 218000 | NA | | [12] |
| Graphene/WS2/ Graphene Heterostructure | 633 nm | 0.1 | 0.03 | NA | Visible | [13] |
| Monolayer WSe2 1.63, DB | 650 nm $V_g = -60V$ $V_{sd} = 2V$ | $1.8\times10^{5}$ | NA | 23 ms | Visible | [14] |
| Graphene/MoS2 | 650 nm | $1.2\times10^{7}$ | 15%(IQE) | NA | | [15] |